# A Fault Tolerance Improved Majority Voter for TMR System Architectures

P. BALASUBRAMANIAN*, K. PRASAD#
* School of Computer Engineering
Nanyang Technological University
50 Nanyang Avenue
SINGAPORE 639798
Email: balasubramanian@ntu.edu.sg
# Department of Electrical and Electronic Engineering
Auckland University of Technology
Auckland 1142
NEW ZEALAND
Email: krishnamachar.prasad@aut.ac.nz

*Abstract:* - For digital system designs, triple modular redundancy (TMR), which is a 3-tuple version of N-modular redundancy is widely preferred for many mission-control and safety-critical applications. The TMR scheme involves two-times duplication of the simplex system hardware, with a majority voter ensuring correctness provided at least two out of three copies of the system remain operational. Thus the majority voter plays a pivotal role in ensuring the correct operation of the system. The fundamental assumption implicit in the TMR scheme is that the majority voter does not become faulty, which may not hold well for implementations based on latest technology nodes with dimensions of the order of just tens of nanometers. To overcome the drawbacks of the classical majority voter some new voter designs were put forward in the literature with the aim of enhancing the fault tolerance. However, these voter designs generally ensure the correct system operation in the presence of either a faulty function module or the faulty voter, considered only in isolation. Since multiple faults may no longer be excluded in the nanoelectronics regime, simultaneous fault occurrences on both the function module and the voter should be considered, and the fault tolerance of the voters have to be analyzed under such a scenario. In this context, this article proposes a new fault-tolerant majority voter which is found to be more robust to faults than the existing voters in the presence of faults occurring internally and/or externally to the voter. Moreover, the proposed voter features less power dissipation, delay, and area metrics based on the simulation results obtained by using a 32/28nm CMOS process.

*Key-Words:* - Digital design, Fault modelling, Fault tolerance, TMR, Majority voter, CMOS, Standard cells

## 1 Introduction

Design-for-reliability and fault-tolerant design have been identified as a major challenge for nanoelectronics designers by the Semiconductor Industry Association's technology roadmap [1]. In this backdrop, fault-tolerant design assumes a greater significance in the nanoelectronics regime, where complicated technological issues such as random dopant (atomistic) fluctuations, sub-wavelength lithography, high heat flux, electro-migration, stress-induced variation, hot carrier effects, negative bias temperature instability, electrostatic discharge, line edge and line width roughness, process-induced defects, and metrology and manufacturing defects [2] [3] impact the manufacturing process. Although some of these technological issues existed at earlier technology nodes, they were less pronounced and were relatively easily dealt with. This is not the case at more recent technology nodes, where these issues tend to pose serious reliability problems for the design of fault-tolerant systems related to safety-intensive applications such as space, aerospace, nuclear, defense, security, power, financial, medical, industrial control and automation, and global positioning and navigation systems.

Fault tolerance basically means guaranteeing the correct operation despite a fault occurrence, and thus signifies higher reliability. Occurrence of a fault internal or external to a function module[1] should not affect the actual output of that function module. If this is ensured, then the fault occurring internally or externally is said to be masked or successfully hidden from being observed by the outside world. On the contrary, if the fault is not masked, it would affect the desired output expected from a function module causing an erroneous output to be produced instead of producing the correct output. Hence, faults which do not cause an error are said to be masked (concealed). On the other hand, faults that result in an error are said to be exposed (revealed). In short, the manifestation of a fault is construed to be an error [4].

Faults can be labelled as transient, intermittent or permanent [4]. Transient/temporary faults are also called as soft errors because they are correctable [5] – [8]. At the logic level, soft errors tend to get manifested as single-event effects [5]. Single-event transients (SETs), which occur due to high-energy particle strikes, might cause a bit-flip at a gate output node or in interconnects formed between logic elements. An SET possessing sufficient amplitude and duration may be captured by a state-holding element in the system stage and subsequently latched, resulting in an error called as single-event upset (SEU) [9]. SEU could also occur when a radiation phenomenon happens to directly flip the binary data output of a register or a memory element which immediately causes an error [10]. SEUs tend to affect data processing in the successive system stage due to permitting computation with erroneous data. However, the manifestation of a transient fault as an error depends upon the electrical, logical, and timing masking of the design [9] [11]. Transient faults can be overcome through radiation hardening of underlying combinational and sequential logic and memory elements by employing redundancy [4] [28].

Intermittent faults [12] refer to those which are activated during certain times and are deactivated during other times, i.e. they occur randomly and might become permanent, for example, a loose electrical connection. Permanent faults are those which imply a physical defect or hardware failure, such as device shorts or opens, broken interconnect etc. which demand repair or replacement.

Permanent faults are generally modelled using stuck-at faults [4] [13]. There are two kinds of stuck-at faults viz. stuck-at-1 (abbreviated as, *s-a*-1) and stuck-at-0 (abbreviated as, *s-a*-0). As the names imply, these faults specify that gate output nodes or interconnects might remain stuck-at the logic high state (i.e. binary 1) or stuck-at the logic low state (i.e. binary 0).There are two kinds of stuck-at faults: single stuck-at fault and multiple stuck-at faults. Single stuck-at fault presumes that a function module contains only one fault. The single stuck-at fault model may not suffice for nanoelectronics digital designs, which encounter greater variability and reliability issues, and so the usage of the multiple faults model is deemed more appropriate [14] [15].

The multiple stuck-at faults model acknowledges that two or more faults can occur at the same time in a function module. Moreover, the multiple stuck-at faults is classified as unidirectional and bidirectional: unidirectional, if all the stuck-at faults are of the same kind (i.e. *s-a*-0 or *s-a*-1); and bidirectional, if the stuck-at faults are different (i.e. *s-a*-0 and *s-a*-1 can co-exist). In this work, without loss of generality, potential transient and permanent faults that may possibly occur shall be represented using the notations 0→1 fault and 1→0 fault, introduced by Pierce in [16]. These notations are elegant in the sense that they can be used to concurrently model both transient as well as permanent faults. 0→1 and 1→0 faults could imply bit-flips due to potential SETs signifying temporary faults, while in the context of permanent faults, 0→1 and 1→0 faults would indicate *s-a*-1 and *s-a*-0 faults respectively.

Radiation hardening by design is widely used to mitigate SETs and SEUs. With respect to radiation hardening by design, both circuit level and system level solutions exist. In the case of ASICs, one of the common circuit level solutions is to custom-develop radiation-tolerant cells which are meant for use in an ASIC-based design synthesis environment.

---

[1] In this article, the term 'function module' is generically used to specify any circuit or system.

However, radiation-tolerant cell designs normally use extra transistors, adopt transistor sizing, and add extra capacitive loads to the output [17] – [20], and hence they potentially occupy more area, consume more power and may be slower in comparison with conventional standard cell libraries. The other viable alternative for designers is to adopt a well-established module level solution such as triple modular redundancy (TMR) [21] [29], which forms a subset of the N-modular redundancy scheme, where three identical function modules are used and a voter[2] is used to produce a majority vote based on the outputs of correctly operating function modules.

The rest of this article is organized as follows. The fundamental TMR scheme is explained in Section 2, and its reliability is compared with that of a simplex system (i.e. non-redundant system) for various module reliabilities. The existing and proposed majority voter designs are presented and described in Section 3, and their fault tolerance are analyzed by considering single and multiple faults occurring internally and/or externally through a simple probabilistic fault analysis metric, proposed in this work. In Section 4, the design parameters viz. power, delay, and area of the different majority voters are estimated using a 32/28nm CMOS process and their respective fault tolerance are also tabulated. Finally, Section 5 concludes this article.

## 2 TMR Scheme

TMR is a generic method, which can be applied for combinational logic, sequential logic, memory cells, and routing elements, individually or in combinations in a digital design. Critics of TMR often point to the excess hardware overhead (about 200%) incurred. To minimize the hardware overhead, approaches for selective insertion of TMR have been proposed in the literature [22] – [25]. Selective application of TMR entails identification of critical circuit portions where TMR can be applied, and non-critical circuit portions where TMR may not be applied. Although not all errors tend to get eliminated through selective TMR insertion, the overall error rate however gets reduced [26]. Selective TMR introduction might serve as a feasible solution to alleviate the overheads of full TMR, especially for applications where weight, cost, and performance also matter besides fault tolerance, such as medical, mobile and portable electronics, and wearable electronics for military purposes. However for mission-critical systems, where reliability is paramount over cost, full TMR is preferred and has been chosen for many space and aerospace applications right from the design of Saturn V Launch Vehicle Digital Computer [27] to the in-flight system design for the Mars Mission [28], and potentially even beyond.

TMR, which forms a subset of N-modular redundancy, requires two-times duplication of a function module and the three identical function modules are joined through a voting element as shown in Figure 1. In Figure 1, function modules 2 and 3 are basically copies of the function module 1. X, Y and Z represent the corresponding (equivalent) outputs of function modules 1, 2 and 3, which form the primary inputs to the voter, whose output is labelled as V. If any arbitrary function module becomes faulty[3], the TMR system will still continue to operate correctly on account of the Boolean majority, which is established by the voter through (1). In (1), product implies logical conjunction, and sum implies logical disjunction.

$$V = XYZ + XY + YZ + XZ = XY + YZ + XZ \qquad (1)$$

Equation (1) inherently assumes that the voter is perfect, i.e. the voter is not faulty. A perfect voter is associated with the ideal reliability value of $R_V$[4] = 1. Under this consideration, the reliability of the TMR system ($R_{TMR}$) is expressed by (2), where the non-faulty state of a function module is represented by $R_M$, and its faulty state is denoted by $(1 – R_M)$. Since the function modules are identical, their reliabilities may also be assumed to be equal. The first term on the right hand side of (2) represents the condition when all the function modules are operating correctly, and the second term on the right hand side of (2) indicates a single function module fault, with the remaining function modules operating correctly.

$$R_{TMR} = R_M^3 + 3(1 – R_M) R_M^2 \qquad (2)$$

The reliability of the simplex system containing just one function module as opposed to three equivalent function modules in a TMR architecture is specified as $R_{Simplex} = R_M$. Figure 2 shows a plot of module reliability (X-axis) versus corresponding system reliabilities (Y-axis) of simplex and TMR systems. It is clear from Figure 2 that up till $R_M < 0.5$, the simplex system is more reliable than the TMR system, although not being fault-tolerant. $R_M = 0.5$ indicates the scenario when the reliabilities of the simplex and TMR systems become equal. However, assuming $R_M$ values to be less than 0.5 is very conservative in a practical scenario as usually function modules tend to have higher reliabilities close to 0.9 [4]. Hence for $R_M$ values greater than 0.5, it is evident from Figure 2 that the TMR system steadily outperforms the simplex system in terms of reliability besides being more fault-tolerant since the simplex system might become a single point-of-failure [4] [30] during critical fault occurrences.

---

[2] Voter/majority voter in this paper, by default, refers to the 2-of-3 majority voter used in TMR circuit/system architectures.

[3] The faulty state of a function module may also imply its catastrophic failure state.

[4] The notation R is used to represent the reliability in this paper. Reliability is akin to probability and is a function of time (*t*). It is implicit in this paper that R = R (*t*) while referring to module or system reliabilities.

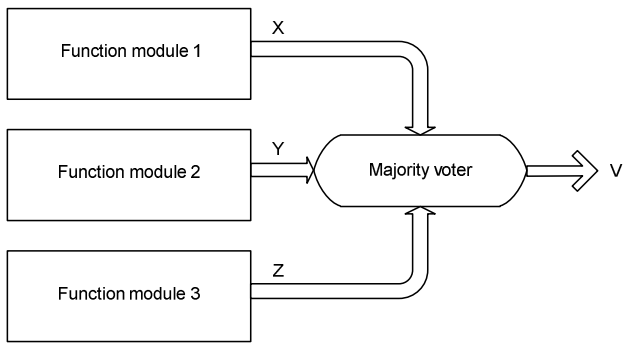

Fig 1. Block diagram of the TMR scheme

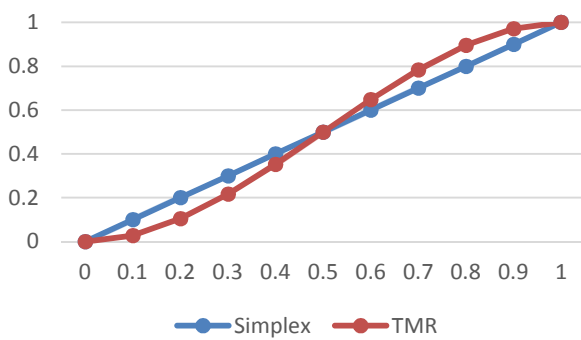

Fig 2. Comparison of reliabilities of simplex and TMR systems (in y-axis) versus module reliability (in x-axis)

# 3 Majority Voters – Designs and Fault Tolerance Analysis

Provided only one function module becomes faulty/fails out of three identical function modules in the TMR scheme, the majority voter is capable of successfully masking a single fault/failure from being noticed by the external environment and also manages to keep the entire system operational. However, it is generally presumed that no fault(s) can occur within the voter, which implies that the voter is assumed to be perfect. If this default assumption is challenged, the TMR system may meet with partial failure, i.e. producing correct outputs for only a subset of the given inputs, or worst might result in complete breakdown. It is important to note that an imperfect or faulty voter may wrongly indicate a system failure when the majority of the function modules are operating correctly, or may erroneously indicate the correct system operation when multiple function modules have indeed become faulty. Hence, besides considering the faulty conditions of function modules and the voter separately, faulty conditions of function modules and the voter also have to be considered simultaneously in order to exhaustively evaluate the fault tolerance capability of the voters. In this section, a number of voter circuits is presented and the possible scenarios for faulty/non-faulty conditions of the function module(s) vis-à-vis a perfect or imperfect voter behavior are illustrated to comprehensively evaluate the fault tolerance of different voter designs.

## 3.1 Classical/Conventional Majority Voter – Fault Tolerance Analysis

The classical voter [21] [29], shown in Figure 3, consists of three 2-input AND gates in the first level and a 3-input OR gate in the second level, which synthesizes (1). This voter shall be referred by the acronym, Classical_MV, for brevity. The acronym 'MV' expands as 'Majority Voter' and shall be used in conjunction with the acronyms of various majority voters in this article. X, Y and Z represent the primary voter inputs, which signify the equivalent outputs of preceding and identical function modules. V represents the voter's output which synthesizes (1). Note that the input and output labels viz. X, Y, Z, and V shall be uniformly maintained throughout this article for all the majority voter designs, and they shall not be repeated further.

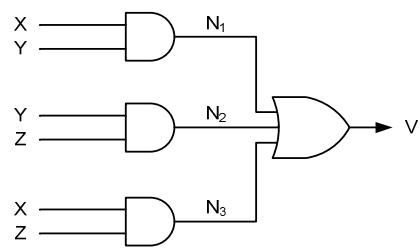

Fig 3. Classical majority voter

The labels $N_1$, $N_2$, $N_3$ of the Classical_MV represent interconnects/internal output nodes of the first-level AND gates in Figure 3. Ideally, the function modules' outputs viz. X, Y and Z which serve as the primary voter inputs are either 0s or 1s, which are specified as the 'no function module fault' conditions in Tables 1 to 4. Note that any other input combination of X, Y and Z signifies 'single/multiple function module faults', as mentioned in Tables 1 to 4. For example, when X, Y and Z are 0, 0 and 1, when they should have been all 0, it is indicative of a single function module fault. On the other hand, when X, Y and Z are 0, 0 and 1 when they are expected to have been 1, it represents multiple function module faults.

Table 1, shown in Appendix A, comprehensively depicts the truth-table of the majority voter shown in Figure 3 and also specifies the potential single/multiple faults that may occur internally within the voter in conjunction with faulty/fault-free function module outputs, and further, captures their subsequent impact on the voter's output. Table 1 additionally serves as a proxy for fault-injection and fault analysis. Hence the proposed 'truth-cum-fault enumeration table' (Tables 1, 2, 3 and 4 of this work) which helps to perform the fault tolerance analysis forms an important contribution of this work. The binary bits shown in blackened boxes under the column 'Internal voter outputs' in Table 1 represents the correct values of internal nodes of the voter viz. $N_1$, $N_2$, and $N_3$ for the applied primary inputs, and they signify the absence of any internal fault occurrence within the voter. The type of fault occurrence viz. 0→1 fault or 1→0 fault on internal nets $N_1$, $N_2$, and $N_3$ is shown annotated in Table 1.

As seen in Table 1, there are many instances when the Classical_MV produces the correct output despite single or multiple function module faults/failures, and there are also a number of instances when the Classical_MV produces an erroneous output. For example, when the primary voter inputs are 111, expected internal outputs are 111 and the voter primary output is expected to be 1. Given this primary input combination, supposing internal output $N_1$, $N_2$ or $N_3$ experiences a single 1→0 fault, this does not affect the primary output and the voter continues to maintain the correct state by masking the internal fault. However, for the same primary input combination, considering the pessimistic case of intermediate nodes $N_1$, $N_2$ and $N_3$ all subject to a 1→0 fault (multiple faults), the voter produces an erroneous output of 0.

To quantitatively evaluate the effect of probable internal voter fault(s) on the voter's primary output, subject to a simultaneous consideration of faulty/fault-free condition of the function modules outputs, a probability-based fault metric viz. the fault masking ratio is proposed and is defined as follows:

- Fault Masking Ratio (FMR) – Specified as the ratio of total number of correct voter output states in the presence of internal and/or external faults, which are masked, divided by the total number of potential internal and/or external fault occurrences

Uniform primary inputs distribution is considered throughout this work to simplify the fault tolerance analysis. Nonetheless, the definition of FMR can be modified to suit a practical scenario by expressing it as the ratio of total number of correct voter output states divided by the total number of likely internal and/or external faults corresponding to the applied primary inputs. FMR is in fact a measure of robustness against potential fault occurrences. From the definition given, it may be understood that FMR has to be high (ideally 1) to achieve good (absolute) fault tolerance. In general, if there is a possibility for $p$ faults to occur, and if $q$ out of $p$ faults are successfully masked from being observed by the outside world, the maximum number of faults that would potentially be exposed to the outside world would be given by ($p - q$). Thus, (1 – FMR) would numerically signify the extent of fault exposure, which has to be low (ideally 0).

For evaluation, FMR pertaining to single and/or multiple faults shall henceforth be denoted by $FMR^{S/MF}$. Referring to Table 1, FMR for the Classical_MV is estimated to be 0.4286, as per the definition. Since the Classical_MV tolerates less than 50% of internal and/or external fault(s), it cannot be labelled as a good fault-tolerant design. This emphasizes the need for a voter design with improved fault tolerance.

### 3.2 Kshirsagar and Patrikar Majority Voter – Fault Tolerance Analysis

The priority encoding based voter proposed by Kshirsagar and Patrikar [31], henceforth identified as the KP_MV, is shown in Figure 4. Two 2-input XOR gates, a priority encoder that consists of an inverter and a 2-input AND gate as shown within the combinational cloud in dotted lines, and a 2:1 multiplexer (MUX) constitute the KP_MV circuit. There are four internal nodes – $N_1$, $N_2$, $N_3$ and P; and these present themselves as candidates for modelling of single/multiple internal faults.

Table 2, shown in Appendix B, portrays a partial truth-cum-fault enumeration of the KP_MV capturing the effect of just a single internal fault on the voter output, subject to single/multiple/no function module faults/failures. As in Table 1, the correct values of intermediate outputs are represented by the binary values shown in blackened boxes under the column 'Internal voter outputs'. The remaining intermediate output values reflect the incorrect binary states due to the presence of only a single fault. When both single and multiple faults which might occur inside the voter are considered in conjunction with single/multiple/no function module faults/failures, Table 2 will comprise a total of 128 listings, which would indeed be numerous to mention here. Hence, only single internal faults are enumerated bit-wise in Table 2 along with an annotation of the type of fault occurrence. Enumeration of multiple faults within the voter can be done in a similar manner as discussed for the Classical_MV, and this is left to the reader.

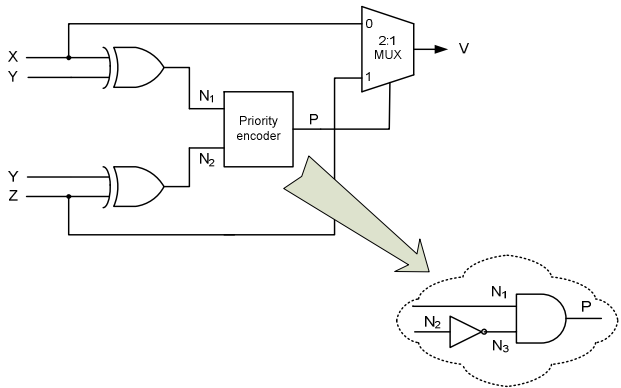

Fig 4. Kshirsagar and Patrikar majority voter

The KP_MV is guaranteed to be fault-proof only in the presence of a single fault occurring internally or externally. However, when a function module becomes faulty/fails and the voter also develops an internal fault, the KP_MV may or may not be fault-tolerant as can be seen in Table 2. For example, when the voter inputs are 0, 1 and 1, and simultaneously if any of the internal nodes becomes faulty, the voter output would be corrupted. When the voter inputs are 1, 0 and 1, and even if any internal node might become faulty, the KP_MV tends to mask the faults. On the other hand, when multiple internal faults occur within the KP_MV, it may cease to be fault-tolerant. Let us now consider two cases for illustration.

- When the primary inputs X, Y and Z are all 1's, internal outputs $N_1$, $N_2$, $N_3$ and P would attain binary values of 0, 0, 1 and 0 respectively and the voter output would correctly evaluate to 1. Under this condition, if any of $N_1$, $N_2$, $N_3$ and P becomes faulty, the voter retains the correct output of 1, thereby the internal fault is successfully masked
- With the primary inputs X, Y and Z now being 1, 1 and 0 respectively, internal outputs $N_1$, $N_2$, $N_3$ and P would attain binary values of 0, 1, 0 and 0 and the voter output would evaluate to 1 since the majority of the inputs is 1. Under this scenario, if the intermediate node P experiences a fault, the

voter would tend to produce an erroneous output of 0, thereby violating the majority convention

From Table 2, and based on the exhaustive consideration of single/multiple internal faults occurring within the voter in conjunction with the faulty (failure)/non-faulty (non-failure) states of the function modules, the FMR of the KP_MV is calculated to be 0.7083. Comparing the FMRs of KP_MV and the Classical_MV, it is clear that the former is more fault-tolerant than the latter by 65.3%. Nevertheless, the KP_MV tends to expose roughly 30% of the faults to the external environment and also features more number of gates, which results in degradation of the design metrics, as substantiated in Section 4.

### 3.3 Ban and Naviner Majority Voter – Fault Tolerance Analysis

Ban and Naviner [32] presented a voter circuit portrayed by Figure 5 which shall henceforth be referred to as BN_MV for brevity. The BN_MV consists of just two gates – a 2-input XOR gate and a 2:1 MUX. Primary inputs X and Y of the voter are XORed and given as the select input for the 2:1 MUX. If the select input is 0, then input Y will be selected and its value will be forwarded to the voter output V. However if the select input is 1, the voter input Z will be reflected on the voter's output. N represents the internal node in Figure 5.

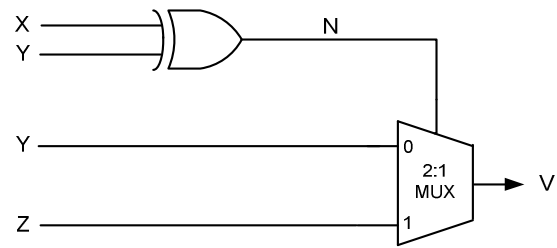

Fig 5. Ban and Naviner majority voter

The BN_MV, similar to the KP_MV, is guaranteed to be fault-proof only when a single fault occurs internally or externally. When a function module fails at random and the voter also develops an internal fault simultaneously, the BN_MV may cease to be fault-tolerant. This is clarified through Table 3, given as Appendix C, which captures all the potential faults that might occur with respect to the internal node N, corresponding to single/multiple/no function module faults. Let us now consider two sample cases to ascertain when the BN_MV tends to be fault-tolerant and when it ceases to be so.

- When the primary inputs X, Y and Z are 0, 1 and 1, the intermediate node N evaluates to 1 and the voter's primary output V also evaluates to 1, since Z is reflected as the voter output. At this juncture, if N is subject to a 1→0 fault, the primary input Y will be selected by the 2:1 MUX and the value of Y which is the same as that of Z will be reflected as the output. Thus, in this case, the BN_MV maintains the correct operation despite an internal and external fault occurrence
- On the other hand, if X, Y and Z are 1, 0 and 1, N is computed as 1 which implies that input Z is selected by the 2:1 MUX and the voter outputs the value of Z which is 1. Given this input scenario, supposing N experiences a 1→0 fault, primary input Y will be selected by the MUX resulting in the voter's output V getting corrupted

Fault tolerance analysis of the BN_MV can be performed based on the data given in Table 3. As in Tables 1 and 2, the correct values of intermediate outputs are indicated by the binary values shown in blackened boxes under the column 'Internal voter outputs'. The remaining internal output values reflect the incorrect binary states due to fault occurrence(s). From Table 3, $FMR^{S/MF}$ of the BN_MV is calculated to be 0.5. This implies the BN_MV allows 50% of the internal and/or external faults to corrupt the primary output. In comparison with the conventional voter, the BN_MV is more fault-tolerant by 16.7%. But compared to the KP_MV, the BN_MV tolerates 29.4% less faults though it has a more compact physical realization. This is substantiated by the design metrics given in Section 4.

### 3.4 Proposed Majority Voter – Design and Fault Tolerance Analysis

The proposed majority voter (Proposed_MV) is shown in Figure 6, which consists of only two gates viz. G1, which is a 2-input OR gate, and G2, which is a complex gate that implements the Boolean function $V = MZ + XY + YZ$, where $M = X + Y$ is the internal output. The Proposed_MV is identical to the BN_MV in that it too features a single internal node. However the Proposed_MV helps to pave the way for improved resilience to potential internal and/or external fault(s).

Table 4, given in Appendix D, captures the truth-cum-fault enumerations of the Proposed_MV. As in the previous Tables, the correct values of intermediate outputs are indicated by the binary values shown in blackened boxes under the column 'Internal voter outputs'. The remaining internal output values reflect the incorrect binary states due to fault occurrences. From Table 4, it can be seen that the Proposed_MV copes with all but one instance of single and multiple function module faults/failures despite the occurrence of any internal fault within it. Hence, with the exception of the two cases where the voter inputs could assume binary values of 001 or 101, and an internal fault may also occur simultaneously within the voter, the proposed majority voter is able to mask all other fault scenario(s) that might occur internally and/or externally. Thus the $FMR^{S/MF}$ of Proposed_MV is high and is calculated to be 0.75.

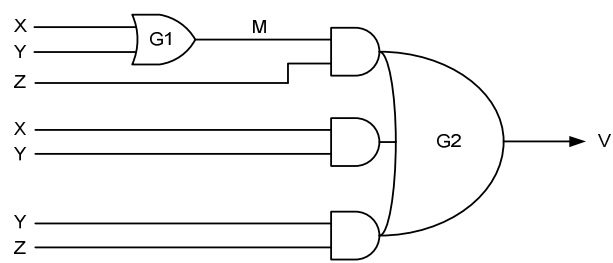

Fig 6. Proposed majority voter

## 4 Simulation Results and Discussion

The different majority voters discussed so far were implemented in semi-custom ASIC design style using the 32/28nm digital standard cell library [33]. The voters were described according to the respective gate-level schematics shown, and their structural integrity was preserved during technology-mapping. This paves the way for a straightforward comparison of the design metrics of different majority voters subsequent to their physical implementation.

Minimum sized gates were chosen uniformly for all the majority voter designs and a typical-case PVT specification was considered with the recommended supply voltage of 1.05V and operating junction temperature of 25°C. Further, wire loads (i.e. parasitic) were included automatically whilst performing the simulations using Synopsys tools. All the primary voter outputs were assigned with fanout-of-4 drive strength. More than 1000 random input vectors corresponding to a diverse sequencing of primary input patterns were applied to the voters at time intervals of 1ns (i.e., 1 GHz) through test benches to capture their switching activities, and the .vcd files thus obtained were subsequently used for average power estimation using Synopsys PrimeTime. The time-based power analysis mode was used to accurately estimate the average power dissipation of the voters. The voters' delay and area were also estimated. The power, delay, area, and FMR of the voters are given in Table 5.

For a combined evaluation of the fault tolerance and design parameters viz. power, delay and area of the different majority voters, a new fault tolerance included figure-of-merit viz. FT-FOM is proposed. FT-FOM is specified as a numerical entity, which is computed as the product of FMR$^{S/MF}$ (in percent) and the figure of merit (FOM), where FOM is specified as the inverse of the product of power, delay, and area (PDAP$^{-1}$). It has already been shown in [34] – [39] that FOM is a useful measure to quantify the physical attributes of a digital design. Since it is desirable to minimize the power, delay, and area metrics, a lower PDAP value and thus a higher FOM are desirable. Moreover, it is desirable to maximize the FMR for achieving enhanced fault tolerance. Hence, a high value of FT-FOM can be considered to be a good indicator of fault tolerance and design performance simultaneously. The calculated FT-FOM values are portrayed in Figure 7.

Table 5. Average power dissipation, maximum propagation delay, area occupancy, and FMR of different majority voters

| Type of voter | Power (µW) | Delay (ns) | Area (µm$^2$) | FMR (%) |
|---|---|---|---|---|
| Classical_MV | 3.52 | 0.13 | 8.39 | 42.86 |
| KP_MV | 6.29 | 0.30 | 15.25 | 70.83 |
| BN_MV | 3.49 | 0.22 | 7.62 | 50 |
| Proposed_MV | 1.88 | 0.17 | 5.34 | 75 |

It can be seen in Figure 7 that the KP_MV has the least FT-FOM as a direct consequence of its higher power dissipation, more propagation delay, and large Silicon area occupancy. The propagation delay of the KP_MV is high as it features more number of logic levels in comparison with the other voters. Since the KP_MV also has more number of logic elements than the other voters, it occupies more area and consequently dissipates more power. Although the FMR of the KP_MV is better than other voter designs and is only less than the FMR of the proposed voter by 5.6%, its FT-FOM is the least among all the voter designs. Though the Classical_MV has the least FMR amongst all the voters, its FT-FOM is indeed greater than the FT-FOM of KP_MV by 3.5×. This is because the Classical_MV has optimized design metrics compared to the KP_MV. The BN_MV, on the other hand, has an enhanced FMR of 14.3% than the Classical_MV, but the latter has an improved FT-FOM of 30.6% compared to the former. On account of less area occupancy, less power dissipation, less propagation delay, and enhanced FMR, the Proposed_MV reports significantly higher FT-FOM than all the other voter designs viz. Classical_MV, KP_MV and BN_MV by 2.9×, 16.9× and 4.1× respectively. Additionally, the peak power dissipation of the voters was estimated and it was found that the Proposed_MV has the least peak power dissipation of 145.2µW amongst all the other voter designs, with the Classical_MV, KP_MV, and BN_MV reporting high peak power dissipations of 176.5µW, 289.7µW, and 234.7µW respectively. The Proposed_MV requires only 18 transistors for physical implementation in static CMOS style after logic factoring [40] through logic optimization. It was shown in [41] that pre-logic factoring followed by physical synthesis could in fact pave the way for optimization of the design metrics.

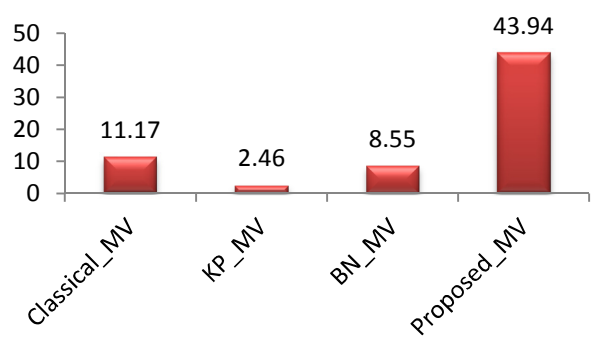

Fig 7. FT-FOM of various voters. X-axis: Voter type; Y-axis: FT-FOM numeric value – a high value of FT-FOM implies high fault masking capability and design efficiency

## 5 Conclusions

The TMR scheme has been widely adopted for numerous mission-control and safety-critical systems applications at both hardware and software levels. At the hardware level, TMR has been predominantly sought after for the fault-tolerant design of ASIC and FPGA based function modules, and all TMR architectures inherently incorporate the majority logic (i.e., the majority voter) to guarantee the correct operation despite a function module fault/failure.

This article has presented different TMR-based majority voter designs, and their realization using a cutting-edge 32/28nm CMOS technology. The fault tolerance viz. fault masking capability of the different majority voter designs has been extensively analyzed by considering the occurrence of single/multiple internal and/or external faults through the truth-cum-fault enumeration table, newly proposed in this work.

It may be noted that the proposed truth-cum-fault enumeration table can be extended to evaluate the fault tolerance property of any digital logic design. Previous related works in the literature have put forward majority voter designs viz. KP_MV and BN_MV, which have better fault tolerance than the classical voter, but these voters are primarily designed to cope with only a single fault occurring internally/externally, and cannot withstand multiple faults/failures that may occur internally and/or externally to the majority voter. With multiple faults becoming common in the era of nanoelectronics, their consideration is deemed important and this has been analyzed at length in this work.

Further, this article has presented a new majority voter design which features improved fault tolerance than the previously proposed majority voter designs with respect to both single and multiple faults occurring internally and/or externally to the voter. A new fault analysis metric, called FMR, was also proposed and used to quantify the fault tolerance of different majority voters when subject to single/multiple fault(s) occurring internally and/or externally, and the proposed majority voter exhibits enhanced FMR than the rest. The estimation of standard design parameters viz. power, delay, and area of the different majority voters has also been done based on a 32/28nm CMOS technology. The fault tolerance analysis and the simulations indicate that the proposed voter achieves superior FMR and quality-of-results (FOM) simultaneously.

APPENDIX A

PART I of Table 1. Truth-cum-fault enumeration table of Classical_MV portraying the effect of internal and/or external faults on the voter output

| Primary voter inputs | | | Internal voter outputs | | | Primary voter output | Voter output state |
|---|---|---|---|---|---|---|---|
| **X** | **Y** | **Z** | **N₁** | **N₂** | **N₃** | **V** | **(Actual/Correct/Error)** |
| *No function module fault/failure* | | | | | | | |
| 0 | 0 | 0 | 0 | 0 | 0 | 0 | Actual |
| | | | 0 | 0 | 1 (0→1) | 1 | Error |
| | | | 0 | 1 (0→1) | 0 | 1 | Error |
| | | | 0 | 1 (0→1) | 1 (0→1) | 1 | Error |
| | | | 1 (0→1) | 0 | 0 | 1 | Error |
| | | | 1 (0→1) | 0 | 1 (0→1) | 1 | Error |
| | | | 1 (0→1) | 1 (0→1) | 0 | 1 | Error |
| | | | 1 (0→1) | 1 (0→1) | 1 (0→1) | 1 | Error |
| *Single/multiple function module faults/failures* | | | | | | | |
| 0 | 0 | 1 | 0 | 0 | 0 | 0 | Actual |
| | | | 0 | 0 | 1 (0→1) | 1 | Error |
| | | | 0 | 1 (0→1) | 0 | 1 | Error |
| | | | 0 | 1 (0→1) | 1 (0→1) | 1 | Error |
| | | | 1 (0→1) | 0 | 0 | 1 | Error |
| | | | 1 (0→1) | 0 | 1 (0→1) | 1 | Error |
| | | | 1 (0→1) | 1 (0→1) | 0 | 1 | Error |
| | | | 1 (0→1) | 1 (0→1) | 1 (0→1) | 1 | Error |
| *Single/multiple function module faults/failures* | | | | | | | |
| 0 | 1 | 0 | 0 | 0 | 0 | 0 | Actual |
| | | | 0 | 0 | 1 (0→1) | 1 | Error |
| | | | 0 | 1 (0→1) | 0 | 1 | Error |
| | | | 0 | 1 (0→1) | 1 (0→1) | 1 | Error |
| | | | 1 (0→1) | 0 | 0 | 1 | Error |
| | | | 1 (0→1) | 0 | 1 (0→1) | 1 | Error |
| | | | 1 (0→1) | 1 (0→1) | 0 | 1 | Error |
| | | | 1 (0→1) | 1 (0→1) | 1 (0→1) | 1 | Error |
| *Single/multiple function module faults/failures* | | | | | | | |
| 0 | 1 | 1 | 0 | 1 | 0 | 1 | Actual |
| | | | 0 | 0 (1→0) | 0 | 0 | Error |
| | | | 0 | 0 (1→0) | 1 (0→1) | 1 | Correct |
| | | | 0 | 1 | 1 (0→1) | 1 | Correct |
| | | | 1 (0→1) | 0 (1→0) | 0 | 1 | Correct |
| | | | 1 (0→1) | 0 (1→0) | 1 (0→1) | 1 | Correct |
| | | | 1 (0→1) | 1 | 0 | 1 | Correct |
| | | | 1 (0→1) | 1 | 1 (0→1) | 1 | Correct |

PART II of Table 1. Truth-cum-fault enumeration table of Classical_MV portraying the effect of internal and/or external faults on the voter output

| Primary voter inputs | | | Internal voter outputs | | | Primary voter output | Voter output state |
|---|---|---|---|---|---|---|---|
| X | Y | Z | $N_1$ | $N_2$ | $N_3$ | V | (Actual/Correct/Error) |
| *Single/multiple function module faults/failures* | | | | | | | |
| 1 | 0 | 0 | 0 | 0 | 0 | 0 | Actual |
| | | | 0 | 0 | 1 (0→1) | 1 | Error |
| | | | 0 | 1 (0→1) | 0 | 1 | Error |
| | | | 0 | 1 (0→1) | 1 (0→1) | 1 | Error |
| | | | 1 (0→1) | 0 | 0 | 1 | Error |
| | | | 1 (0→1) | 0 | 1 (0→1) | 1 | Error |
| | | | 1 (0→1) | 1 (0→1) | 0 | 1 | Error |
| | | | 1 (0→1) | 1 (0→1) | 1 (0→1) | 1 | Error |
| *Single/multiple function module faults/failures* | | | | | | | |
| 1 | 0 | 1 | 0 | 0 | 1 | 1 | Actual |
| | | | 0 | 0 | 0 (1→0) | 0 | Error |
| | | | 0 | 1 (0→1) | 0 (1→0) | 1 | Correct |
| | | | 0 | 1 (0→1) | 1 | 1 | Correct |
| | | | 1 (0→1) | 0 | 0 (1→0) | 1 | Correct |
| | | | 1 (0→1) | 0 | 1 | 1 | Correct |
| | | | 1 (0→1) | 1 (0→1) | 0 (1→0) | 1 | Correct |
| | | | 1 (0→1) | 1 (0→1) | 1 | 1 | Correct |
| *Single/multiple function module faults/failures* | | | | | | | |
| 1 | 1 | 0 | 1 | 0 | 0 | 1 | Actual |
| | | | 0 (1→0) | 0 | 0 | 0 | Error |
| | | | 0 (1→0) | 0 | 1 (0→1) | 1 | Correct |
| | | | 0 (1→0) | 1 (0→1) | 0 | 1 | Correct |
| | | | 0 (1→0) | 1 (0→1) | 1 (0→1) | 1 | Correct |
| | | | 1 | 0 | 1 (0→1) | 1 | Correct |
| | | | 1 | 1 (0→1) | 0 | 1 | Correct |
| | | | 1 | 1 (0→1) | 1 (0→1) | 1 | Correct |
| *No function module fault/failure* | | | | | | | |
| 1 | 1 | 1 | 1 | 1 | 1 | 1 | Actual |
| | | | 0 (1→0) | 0 (1→0) | 0 (1→0) | 0 | Error |
| | | | 0 (1→0) | 0 (1→0) | 1 | 1 | Correct |
| | | | 0 (1→0) | 1 | 0 (1→0) | 1 | Correct |
| | | | 0 (1→0) | 1 | 1 | 1 | Correct |
| | | | 1 | 0 (1→0) | 0 (1→0) | 1 | Correct |
| | | | 1 | 0 (1→0) | 1 | 1 | Correct |
| | | | 1 | 1 | 0 (1→0) | 1 | Correct |

APPENDIX B

PART I of Table 2. Partial truth-cum-fault enumeration table of KP_MV, highlighting the effect of single internal fault on the voter output in the presence of single/multiple/no function module faults/failure

| Primary voter inputs | | | Internal voter outputs | | | | Primary voter output | Voter output state |
|---|---|---|---|---|---|---|---|---|
| X | Y | Z | $N_1$ | $N_2$ | $N_3$ | P | V | (Actual/Correct/Error) |
| | | | *No function module fault/failure* | | | | | |
| 0 | 0 | 0 | 0 | 0 | 1 | 0 | 0 | Actual |
| | | | 0 | 0 | 0 (1→0) | 0 | 0 | Correct |
| | | | 0 | 0 | 1 | 1 (0→1) | 0 | Correct |
| | | | 0 | 1 (0→1) | 0 | 0 | 0 | Correct |
| | | | 1 (0→1) | 0 | 1 | 1 | 0 | Correct |
| | | | *Single/multiple function module faults/failures* | | | | | |
| 0 | 0 | 1 | 0 | 1 | 0 | 0 | 0 | Actual |
| | | | 0 | 1 | 0 | 1 (0→1) | 1 | Error |
| | | | 0 | 1 | 1 (0→1) | 0 | 0 | Correct |
| | | | 0 | 0 (1→0) | 1 | 0 | 0 | Correct |
| | | | 1 (0→1) | 1 | 0 | 0 | 0 | Correct |
| | | | *Single/multiple function module faults/failures* | | | | | |
| 0 | 1 | 0 | 1 | 1 | 0 | 0 | 0 | Actual |
| | | | 1 | 1 | 0 | 1 (0→1) | 0 | Correct |
| | | | 1 | 1 | 1 (0→1) | 1 | 0 | Correct |
| | | | 0 (1→0) | 1 | 0 | 0 | 0 | Correct |
| | | | 1 | 0 (1→0) | 1 | 1 | 0 | Correct |
| | | | *Single/multiple function module faults/failures* | | | | | |
| 0 | 1 | 1 | 1 | 0 | 1 | 1 | 1 | Actual |
| | | | 1 | 0 | 0 (1→0) | 0 | 0 | Error |
| | | | 1 | 0 | 1 | 0 (1→0) | 0 | Error |
| | | | 1 | 1 (0→1) | 0 | 0 | 0 | Error |
| | | | 0 (1→0) | 0 | 1 | 0 | 0 | Error |
| | | | *Single/multiple function module faults/failures* | | | | | |
| 1 | 0 | 0 | 1 | 0 | 1 | 1 | 0 | Actual |
| | | | 1 | 0 | 0 (1→0) | 0 | 1 | Error |
| | | | 1 | 0 | 1 | 0 (1→0) | 1 | Error |
| | | | 0 (1→0) | 0 | 1 | 0 | 1 | Error |
| | | | 1 | 1 (0→1) | 0 | 0 | 1 | Error |
| | | | *Single/multiple function module faults/failures* | | | | | |
| 1 | 0 | 1 | 1 | 1 | 0 | 0 | 1 | Actual |
| | | | 1 | 1 | 0 | 1 (0→1) | 1 | Correct |
| | | | 1 | 1 | 1 (0→1) | 1 | 1 | Correct |
| | | | 1 | 0 (1→0) | 1 | 1 | 1 | Correct |
| | | | 0 (1→0) | 1 | 0 | 0 | 1 | Correct |

PART II of Table 2. Partial truth-cum-fault enumeration table of KP_MV, highlighting the effect of single internal fault on the voter output in the presence of single/multiple/no function module faults/failure

| Primary voter inputs | | | Internal voter outputs | | | | Primary voter output | Voter output state |
|---|---|---|---|---|---|---|---|---|
| X | Y | Z | $N_1$ | $N_2$ | $N_3$ | P | V | (Actual/Correct/Error) |
| *Single/multiple function module faults/failures* | | | | | | | | |
| 1 | 1 | 0 | 0 | 1 | 0 | 0 | 1 | Actual |
| | | | 0 | 1 | 0 | 1 (0→1) | 0 | Error |
| | | | 0 | 1 | 1 (0→1) | 0 | 1 | Correct |
| | | | 0 | 0 (1→0) | 1 | 0 | 1 | Correct |
| | | | 1 (0→1) | 1 | 0 | 0 | 1 | Correct |
| *No function module fault/failure* | | | | | | | | |
| 1 | 1 | 1 | 0 | 0 | 1 | 0 | 1 | Actual |
| | | | 0 | 0 | 0 (1→0) | 0 | 1 | Correct |
| | | | 0 | 0 | 1 | 1 (0→1) | 1 | Correct |
| | | | 0 | 1 (0→1) | 0 | 0 | 1 | Correct |
| | | | 1 (0→1) | 0 | 1 | 1 | 1 | Correct |

APPENDIX C

PART I of Table 3. Truth-cum-fault enumeration table of BN_MV, capturing the effect of internal and/or external faults on the voter output

| Primary voter inputs | | | Internal voter output | Primary voter output | Voter output state |
|---|---|---|---|---|---|
| X | Y | Z | N | V | (Actual/Correct/Error) |
| *No function module fault/failure* | | | | | |
| 0 | 0 | 0 | 0 | 0 | Actual |
| | | | 1 (0→1) | 0 | Correct |
| *Single/multiple function module faults/failures* | | | | | |
| 0 | 0 | 1 | 0 | 0 | Actual |
| | | | 1 (0→1) | 1 | Error |
| *Single/multiple function module faults/failures* | | | | | |
| 0 | 1 | 0 | 1 | 0 | Actual |
| | | | 0 (1→0) | 1 | Error |
| *Single/multiple function module faults/failures* | | | | | |
| 0 | 1 | 1 | 1 | 1 | Actual |
| | | | 0 (1→0) | 1 | Correct |
| *Single/multiple function module faults/failures* | | | | | |
| 1 | 0 | 0 | 1 | 0 | Actual |
| | | | 0 (1→0) | 0 | Correct |
| *Single/multiple function module faults/failures* | | | | | |
| 1 | 0 | 1 | 1 | 1 | Actual |
| | | | 0 (1→0) | 0 | Error |

PART II of Table 3. Truth-cum-fault enumeration table of BN_MV, capturing the effect of internal and/or external faults on the voter output

| Primary voter inputs | | | Internal voter output | Primary voter output | Voter output state |
|---|---|---|---|---|---|
| X | Y | Z | N | V | (Actual/Correct/Error) |
| *Single/multiple function module faults/failures* | | | | | |
| 1 | 1 | 0 | 0 | 1 | Actual |
| | | | 1 (0→1) | 0 | Error |
| *No function module fault/failure* | | | | | |
| 1 | 1 | 1 | 0 | 1 | Actual |
| | | | 1 (0→1) | 1 | Correct |

APPENDIX D

Table 4. Truth-cum-fault enumeration table of Proposed_MV showing the effect of internal and/or external faults on the voter output

| Primary voter inputs | | | Internal voter output | Primary voter output | Voter output state |
|---|---|---|---|---|---|
| X | Y | Z | M | V | (Actual/Correct/Error) |
| *No function module fault/failure* | | | | | |
| 0 | 0 | 0 | 0 | 0 | Actual |
| | | | 1 (0→1) | 0 | Correct |
| *Single/multiple function module faults/failures* | | | | | |
| 0 | 0 | 1 | 0 | 0 | Actual |
| | | | 1 (0→1) | 1 | Error |
| *Single/multiple function module faults/failures* | | | | | |
| 0 | 1 | 0 | 1 | 0 | Actual |
| | | | 0 (1→0) | 0 | Correct |
| *Single/multiple function module faults/failures* | | | | | |
| 0 | 1 | 1 | 1 | 1 | Actual |
| | | | 0 (1→0) | 1 | Correct |
| *Single/multiple function module faults/failures* | | | | | |
| 1 | 0 | 0 | 1 | 0 | Actual |
| | | | 0 (1→0) | 0 | Correct |
| *Single/multiple function module faults/failures* | | | | | |
| 1 | 0 | 1 | 1 | 1 | Actual |
| | | | 0 (1→0) | 0 | Error |
| *Single/multiple function module faults/failures* | | | | | |
| 1 | 1 | 0 | 1 | 1 | Actual |
| | | | 0 (1→0) | 1 | Correct |
| *No function module fault/failure* | | | | | |
| 1 | 1 | 1 | 1 | 1 | Actual |
| | | | 0 (1→0) | 1 | Correct |